\documentclass[a4paper, conference]{IEEEtran}

\usepackage{graphicx}
\usepackage{epstopdf}
\usepackage{amsthm}
\usepackage{amsmath}
\usepackage{amsfonts}
\usepackage{amssymb}

\newtheorem{definition}{Definition}

\ifCLASSINFOpdf
\else
\fi
\begin{document}
%
\title{Data Science: Nature and Pitfalls}


\author{Longbing Cao\\
Advanced Analytics Institute, University of Technology Sydney, Australia \\ 

}


%

\makeatletter
\def\ps@headings{%
\def\@oddhead{\mbox{}\scriptsize\rightmark \hfil \thepage}%
\def\@evenhead{\scriptsize\thepage \hfil \leftmark\mbox{}}%
\def\@oddfoot{}%
\def\@evenfoot{}}
\makeatother

\maketitle
\pagestyle{empty}
\thispagestyle{empty}
\IEEEpeerreviewmaketitle

\begin{abstract}
Data science is creating very exciting trends as well as significant controversy. A critical matter for the healthy development of data science in its early stages is to deeply understand the nature of data and data science, and to discuss the various pitfalls. These important issues motivate the discussions in this article.
\end{abstract}

\renewcommand{\IEEEkeywordsname}{Keywords}


%

\IEEEpeerreviewmaketitle

\section{Introduction}
\label{sec:intro}

The era of analytics \cite{Tukey62}, data science \cite{Tukey77}, and big data \cite{Mckinsey11} has driven substantial governmental, industrial, and disciplinary interest, goal and strategy transformation as well as a paradigm shift in research and innovation. This has resulted in  significant new opportunities and prospects becoming available which were not previously possible, while an increasing and overwhelming amount of fanfare and hype has spread across multiple domains, areas and events.

In reviewing the related initiatives, progress, and status of  data science, analytics and big data \cite{cao16-2}, and the diversified discussions about prospects, challenges and directions \cite{Cao16ds}, the controversy caused by the potential conflict of these various elements becomes clear. This review discloses the need for deep discussions about the nature and pitfalls of data science, the clarification of fundamental concepts and myths, and the demonstration of the intrinsic characteristics and opportunities of data science. 

This paper thus focuses on discussing the two fundamental issues in data science: the nature and the pitfalls of data science. Addressing the first highlights the status, intrinsic factors, characteristics, and features of the era of data science and analytics, as well as the challenges and opportunities for new generation innovation, research and disciplinary development. The second summarizes common pitfalls about the concepts of data science, data volume, infrastructure, analytics, and capabilities and roles. Building on the above discussions, the concepts and possible future directions of data science are then presented.

\section{Features of the Data Science Era}
\label{sec:features}

Drawing a picture of the features and  characteristics of the era of data science is critical and challenging. We explore this from the perspective of the transformation and paradigm shift caused by data science, and discuss the core driving forces, as well as the status of a number of typical issues confronting the data science field.

\subsection{The Era's Transformation and Paradigm Shift}
\label{subsec:transformation}
The emergence of the era of data science and analytics can be highlighted by three key indicators: a significant \textit{disciplinary paradigm shift}, \textit{technological transformation}, and \textit{innovative production}. 
\begin{itemize}
\item Disciplinary paradigm shifting: The shifting of data-centric disciplinary paradigms from one to another.
\item Technological transformation: The upgrading of data technology from one generation to another.
\item Innovative production: The innovation of technical and practical data products.
\end{itemize}
These are discussed below.

The disciplinary paradigm shift of data-oriented and data-centric research, innovation and profession can be embodied by such aspects as:
\begin{itemize}
\item From data analysis to data analytics;
\item From descriptive analytics to deep analytics;
\item From data analytics to data science;
\end{itemize}

The disciplinary paradigm shift promotes data-related technological transformation by such means as follows:  
\begin{itemize}
\item From large-scale data to big data;
\item From business operational systems to business analytical systems;
\item From World Wide Web to Wisdom web;
\item From Internet to Internet of Everything (incl. Internet of Things, mobile network, social network);
\end{itemize}

Innovative production in the data and analytics areas can be represented by typical indicators such as the following:
\begin{itemize}
\item From digital economy to data economy; 
\item From closed government to open government;
\item From e-commerce to online business;
\item From telephone to smart phone;
\item From Internet to mobile and social network.
\end{itemize}

\subsection{Data-centric Driving Forces}
\label{subsec:drivers}

The transformation and paradigm shift of data-oriented discipline, technologies and production are driven by core forces including data-enabled opportunities, data-related ubiquitous factors, and various complexities and intelligences embedded in data-oriented production and products.

Ubiquitous data-oriented factors include data, behavior, complexity, intelligence, service, and opportunities.
\begin{itemize}
\item Data is ubiquitous: Involving historical, real-time and future data;
\item Behavior is ubiquitous: Bridging the gaps between the physical world and the data world;
\item Complexity is ubiquitous: The types and extent of complexity differentiate one data system from another;
\item Intelligence is ubiquitous: Diversified intelligences are embedded in a data system;
\item Service is ubiquitous: Data services present in various forms and domains;
\item Opportunities are ubiquitous: Data enables enormous opportunities.
\end{itemize}

Data-enabled opportunities, also called \textit{X-opportunities}, are overwhelming and extend from research, innovation, education, and government to economy. We briefly elaborate on them below.
\begin{itemize}
\item Research opportunities: Inventing data-focused breakthrough theories and technologies;
\item Innovation opportunities: Developing data-based cutting-edge services, systems and tools;
\item Education opportunities: Innovating data-oriented courses and training;
\item Government opportunities: Enabling data-driven government decision-making and objectives;
\item Economic opportunities: Fostering data economy, services, and industrialization;
\item Lifestyle opportunities: Promoting data-enabled smarter living and smarter cities;
\item Entertainment opportunities: Creating data-driven entertainment activities, networks, and societies.
\end{itemize}

A data science problem is a complex system \cite{Mitchel2011,Metasynthetic15} in which comprehensive system complexities, also called \textit{X-complexities} \cite{Cao16ds}, are embedded. These comprise complexities of data characteristics, behavior, domain, society (social complexity), environment, learning, and decision-making.
\begin{itemize}
\item Data complexity: Embodied by such factors as comprehensive data circumstances and characteristics;
\item Behavior complexity: Demonstrated by such aspects as individual and group activities, evolution, utility, impact and change; 
\item Domain complexity: Represented by such aspects as domain factors, processes, norms, policies, knowledge and domain expert engagement in problem-solving;
\item Social complexity: Indicated by such aspects as social networking, community formation and divergence, sentiment, the dissemination of opinion and influence, and other social issues such as trust and security;
\item Environment complexity: Capturing such aspects as contextual factors, interactions with systems, changes, and uncertainty.
\item Learning complexity: Including the development of appropriate methodologies, frameworks, processes, models and algorithms, and theoretical foundation and explanation;
\item Decision complexity: Involving such issues as the methods and forms of deliverables, communications and decision-making actions. 
\end{itemize}

In a complex data science problem, ubiquitous intelligence, also called \textit{X-intelligence} \cite{Cao16ds}, is often demonstrated and has to be incorporated and synergized \cite{Metasynthetic15} in problem-solving processes and systems.
\begin{itemize}
\item Data intelligence: Highlighting the interesting information, insights and stories hidden in data about business problems and driving forces.
\item Behavior intelligence: Demonstrating the  insights of activities, processes, dynamics, impact and trust of individual and group behaviors by human and action-oriented organisms. 
\item Domain intelligence: Domain values and insights emerge from involving domain factors, knowledge, meta-knowledge, and other domain-specific resources.
\item Human intelligence: Contributions made by the empirical knowledge, beliefs, intentions, expectations, critical thinking and imaginary thinking of human individual and group actors.
\item Network intelligence: Intelligence created by the involvement of networks, web, and networking mechanisms in problem comprehension and problem-solving.
\item Organizational intelligence: Insights and contributions created by the involvement of organization-oriented factors, resources, competency and capabilities, maturity, evaluation and dynamics.
\item Social intelligence: Contributions and values generated by the inclusion of social, cultural, and economic factors, norms and regulation.
\item Environmental intelligence: This may be embodied through other intelligences  specific to the underlying domain, organization, society, and actors. 
\end{itemize}

The above data-oriented and data-driven factors, complexities,  intelligences and opportunities constitute the nature and characteristics of data science, and drive the evolution and dynamics of data science problems.

\subsection{Data DNA}
\label{subsec:dna}

In the biological domain, DNA is a molecule that carries  genetic instructions that are uniquely valuable to the biological development, functioning and reproduction of humans and all known living organisms. As a result of data quantification, data is everywhere, and is present in the public Internet, Internet of Things, sensor networks, socio-cultural, economic and geographical repositories and quantified personalized sensors, including mobile, social, living, entertaining and emotional sources. This forms the ``datalogical'' constituent: ``data DNA'', which plays a critical role in data organisms and performs a similar function to biological DNA in living organisms. 

\begin{definition}[Data DNA]
\emph{Data DNA} is the datalogical ``molecule'' of data, consisting of fundamental and generic constituents: entity (E), property (P) and relationship (R). Here ``datalogical'' means that data DNA plays a similar role in data organisms as biological DNA plays in living organisms. \emph{Entity} can be an object, an instance, a human, an organization, a system, or a part of a sub-system of a system. \emph{Property} refers to the attributes that describe an entity. \emph{Relationship} corresponds to (1) entity interactions, and (2) property interactions, including property value interactions. 
\end{definition}

Entity, property and relationship present different characteristics in terms of quantity, type, hierarchy, structure, distribution and organization. A data-intensive application or system is often composed of a large number of diverse entities, each of which has specific properties, and different relationships are embedded within and between properties and entities. From the very lowest level to the very highest level, data DNA presents heterogeneity and hierarchical couplings across levels. On each level, it maintains \textit{consistency} (inheritance of properties and relationships) as well as \textit{variations} (mutations) across entities, properties and relationships, while \textit{personalized characteristics} are supported for each individual entity, property and relationship. 

For a given data, its entities, properties and relationships are instantiated into diverse and domain-specific forms, which carry most of the data ecological and genetic information in data generation, development, functioning, reproduction, and evolution.  
In the data world, \emph{data DNA} is embedded in the whole body of  personal \cite{personaldata11} and non-personal data organisms, and in the generation, development, functioning, management, analysis and use of all data-based applications and systems. 

Data DNA drives the evolution of a data-intensive organism. For example, university data DNA connects the data of students, lecturers, administrative systems, corporate services and operations. The student data DNA further consists of academic, pathway, library access, online access, social media, mobile service, GPS, and Wifi usage data. Such student data DNA is both steady and evolving. 

In complex data, data DNA is embedded within various X-complexities (see detailed discussions in \cite{Cao16ds} and \cite{Metasynthetic15}) and ubiquitous X-intelligence (more details in \cite{Cao16ds} and \cite{Metasynthetic15}) in a data organism. This makes data rich in content, characteristics, semantics and value, but challenging in acquisition, preparation, presentation, analysis and interpretation. 

\subsection{Data Quality}
\label{subsec:quality}

Data science tasks involve roles and follow processes different from more generalized IT projects, since data science and analytics works tend to be  creative, intelligent, exploratory, non-standard, unautomated and personalized, and have the objective of discovering evidence and indicators for decision-making actions. They inevitably involve quality issues such as data validity, veracity, variability and reliability, and social issues such as privacy, security, accountability and trust, which need to be taken into account in data science and analytics. 

\textit{Data quality} is a critical problem in data science and engineering. Given a data science problem, we should not assume that 
\begin{itemize}
\item The data available or given is perfect,
\item The data always generates good outcomes,
\item The outputs (findings) generated are always good and meaningful, and
\item The outcomes can always inform better decisions.
\end{itemize} 

These assumption myths involve the quality of the  data (input), the model, and the outcomes (output), in particular, validity, veracity, variability and reliability. We briefly discuss these aspects below.    

Data and analytics \emph{validity} determines whether a data model, concept, conclusion or measurement is well-founded and corresponds accurately to the data characteristics and real-world facts, making it capable of giving the right answer. 

Similarly, data and analytics \emph{veracity} determines the correctness and accuracy of data and analytics outcomes. Both validity and veracity also need to be checked from the perspectives of data content, representation, design, modeling, experiments, and evaluation.

Data and analytics \emph{variability} is determined by the changing and uncertain nature of data, reflecting  business dynamics (including the problem context and problem-solving purposes), and thus requires the corresponding analytics to adapt to the dynamic nature of the data. Due to the changing nature of data, the need to check the validity, veracity and reliability of the data used and analytics undertaken is thus highly important.
   
Data and analytics \emph{reliability} refers to the consistency, redundancy, repeatability and trust properties of the data used, the analytic models generated, and the outcomes delivered on the data. Reliable data and analytics are not necessarily static. Making data analytics adaptive to the evolving, streaming and dynamic nature of data, business and decision requests is a critical challenge in data science and analytics. 

\begin{figure*}[ht!]
\centerline{\includegraphics[width=0.65\textwidth]{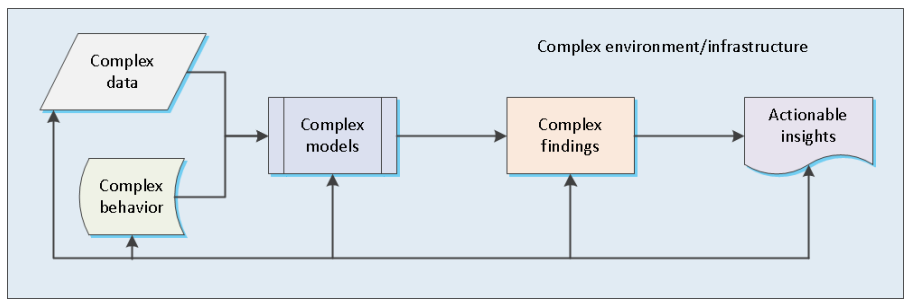}}
\caption{The extreme data challenge.}
\label{fig:five}
\end{figure*}

\subsection{Social Issues}
\label{subsec:socialissues}

Domain-specific data and business are embedded in social contexts and incorporated with social issues. Data science tasks typically involve such social issues as \textit{privacy}, \textit{security}, \textit{accountability} and \textit{trust} on data, modeling and deliverables, which we discuss below.

Data and analytics \emph{privacy} addresses the challenge of collecting, analyzing, disseminating and sharing data and analytics while protecting personally identifiable or other sensitive information and analytics from improper disclosure. Protection technology, regulation and policies are required to balance  protection and appropriate disclosure in the process of data manipulation. 

Data and analytics \emph{security} protects target objects from destructive forces and from the unwanted actions of unauthorized users, including improper use or disclosure, and not only addresses privacy issues but also other aspects beyond privacy, such as software and hardware backup and recovery. Data and analytics security also involves the development of regulating, political or  legal mechanisms and systems to address such issues.

Data and analytics \emph{accountability} refers to an obligation to comply with data privacy and security legislation, and  to report, explain, trace and identify the data manipulated and analytics conducted to maintain the transparency and traceability, liability and warranty of both measurement and results, as well as the efficacy and verifiability of analytics and protection.

Data and analytics \emph{trust} refers to the belief in the reliability, truth or ability of data and analytics to achieve the relevant goals. This involves the development of appropriate technology, social norms, ethical rules, or  legislation to ensure, measure and protect trust in the data and analytics used and confidence in the corresponding outcomes and evaluation of analytics.

\subsection{The Extreme Challenge}
\label{subsec:extrechallenge}

Different types and levels of analytical problems trouble the existing knowledge base, and we are especially challenged by the complex problems in complex data and in complex environments. Our particular focus on data science research and innovation concerns the type of scenario we call an \emph{extreme data challenge} in data science and analytics. The \emph{extreme data challenge as illustrated in Fig. \ref{fig:five}, seeks  to discover and deliver complex knowledge in complex data taking into account complex behavior within a complex environment to achieve actionable insights that will inform and enable  decision action-taking in complex business problems} that cannot be better handled by other methods.

The critical future directions of data science research and innovation in this case are focused on 
\begin{itemize}
\item \textit{Complex data} with complex characteristics and various complexities (as discussed above on data complexity and intelligence; for more, see \cite{Cao16ds} and \cite{Metasynthetic15}),
\item \textit{Complex behaviors} with complex relationships and dynamics (as discussed above on behavior complexity and intelligence; for more, see \cite{Cao16ds} and \cite{Metasynthetic15}),
\item \textit{Complex environment} in which complex data and behaviors are embedded and interacted with (as discussed above on domain-specific, organizational, social and environmental complexities and intelligence; for more, see \cite{Cao16ds} and \cite{Metasynthetic15}), 
\item \textit{Complex models} to address the data and behavior complexities in complex environment (as discussed about learning complexities and decision complexities),
\item \textit{Complex findings} to uncover hidden but technically interesting and business-friendly observations, indicators or evidence, statements or presentations, and 
\item \textit{Actionable insights} to evidence the next best or worst situation and inform the optimal strategies that should be taken to support effective business decision-making (see more discussion on actionability in \cite{Cao10dddm,dddm10}.
\end{itemize}
 
Many real-life problems fall into this level of complexities and challenges, as shown in the extreme data challenge, and they have not been addressed well; for example,
\begin{itemize}
\item Understanding group behaviors  by multiple actors where there are complex interactions and relationships, such as in the manipulation of large-scale cross-capital markets pool  by internationally collaborative investors \cite{Cao2012cba}, each of whom plays a specific role by connecting information from the underlying markets, social media, other financial markets, socio-economic data and policies \cite{noniid14};
\item Predicting local climate change and effect by connecting local, regional and global climate data, geographical data, agricultural data and other information \cite{Faghmous14}. 
\end{itemize}

\section{Disciplinary Development of Data Science}
\label{sec:opp}

In this section, we present a status summary of the disciplinary development of data science by reviewing the development gaps between the potential that data may have and the state-of-the-art capabilities to fulfill such potential, the research map of data science, and the course framework of data science. 

\subsection{Data-to-Capability Development Gaps}
\label{subsec:gaps}

\begin{figure*}[ht!]
\centerline{\includegraphics[width=0.85\textwidth]{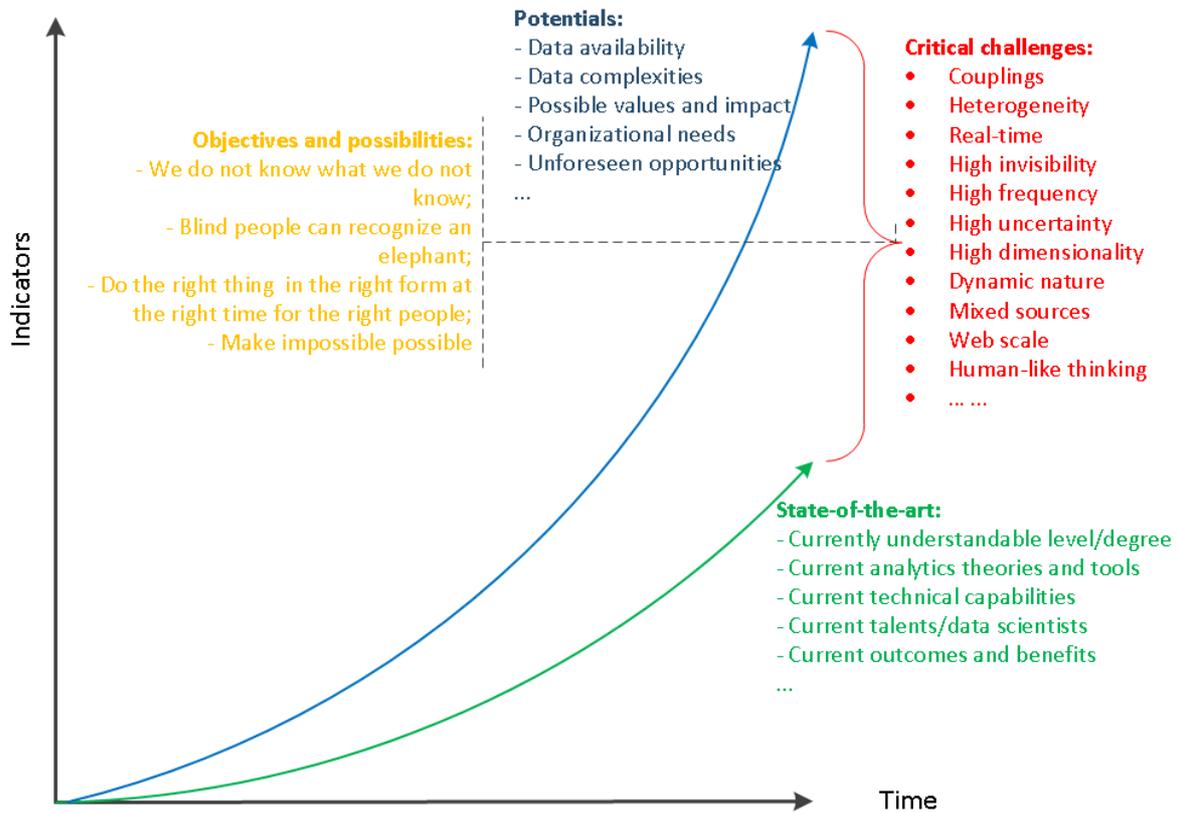}}
\caption{Critical development gaps between data potentials and state-of-the-art capabilities.}
\label{fig:four}
\end{figure*}

The rapid increase in big data unfortunately does not only present opportunities. As discussed in \cite{Cao10dddm} and \cite{dddm10}, there are significant gaps between what we are of and what we are capable of understanding. An  empirical observation of the \emph{data development gaps} between (1) the growth of \textit{data potentials}  and (2) the \textit{state-of-the-art capabilities} is shown in Fig. \ref{fig:four}. Such gaps have increased in the last 10 years and especially recently, due to the imbalance between potential exponential  increase and progressive state-of-the-art capability development. We illustrate several such gaps below. 
\begin{itemize}
\item Gap between data availability and currently understandable data level, scale and degree;
\item Gap between data complexities and currently available analytics theories and tools;
\item Gap between data complexities and currently available technical capabilities;
\item Gap between possible values and impact and currently achievable outcomes and benefits;
\item Gap between organizational needs and currently available talents/data scientists;
\item Gap between potential opportunities and current outcomes and benefits achievable.
\end{itemize}

Such growth gaps are driven by critical challenges for which there is a shortage of effective theories and tools. For example, a typical challenge in complex data concerns intrinsic complex coupling relationships and heterogeneity, forming non-IID data \cite{noniid14},  which cannot be simplified in such a way that they can be handled by classic IID learning theories and systems. Other examples include the real-time learning of large-scale online data, such as learning shopping manipulation and making real-time recommendations on extremely high frequency data in the ``11-11'' shopping seasons launched by Alibaba, or identifying suspects in an extremely imbalanced and multi-source data and environment such as fraud detection in high frequency marketing trading. Other challenges are high invisibility, high frequency, high uncertainty, high dimensionality, dynamic nature, mixed sources, online learning at the web scale, and developing human-like thinking.      

\subsection{Research Map of Data Science}
\label{subsec:advanalytics}

The way to explore the fundamental challenges and innovative opportunities facing big data and data science is to conduct problem-, data-, and goal-driven discovery.
\begin{itemize}
\item \textit{Problem-driven discovery}: This requires understanding the intrinsic nature, characteristics, complexities, and boundaries of the problem, and then analyzing the gaps between the problem complexities and the existing capability set. This gap analysis is critical for original research and breakthrough scientific discovery. 
\item \textit{Goal-driven discovery}: This requires understanding the business, technical and decision goals to be achieved by understanding the problem, and conducting gap analysis of what has been implemented and achieved and what is expected to be achieved.
\item \textit{Data-driven discovery}: This requires understanding the data characteristics, complexities and challenges in data, and the gaps between the nature of a problem and the data capabilities. Due to the limitations of existing data systems,  projection from the underlying physical world where the problem sits to the data world where the problem is datafied may be biased, dishonest, or highly manipulated. As a result, the data does not completely capture the problem and thus cannot create a full picture of of the through any type of data exploration.   
\end{itemize}

There are two ways to explore major research challenges: one is to summarize what concerns the relevant communities, and the other is to scrutinize the potential issues arising from the intrinsic complexities and nature of data science problems as complex systems \cite{Metasynthetic15}, \cite{Metasynthetic15}. Taking the first approach, we can obtain a picture of the main research challenges by summarizing the main topics and issues  in the statistics communities \cite{Chambers93}, \cite{wu97}, \cite{amstatnews15}, informatics and computing communities  \cite{Rudin14}, \cite{Cao16ds}, vendors \cite{Stonebraker13}, government initiatives \cite{USNSF}, \cite{UNpulse}, \cite{CNbd}, \cite{ECbd14}, \cite{UKbd} and research institutions \cite{UTSAAI}, \cite{IAA}, which focus on data science and analytics. The second approach is much more challenging, as it requires us to explore the unknown space of the complexities and comprehensive intelligence in complex data problems. 


Below, we list some of the main challenges confronting the data science community in addressing the big data complexities represented by key topics in the data A-Z (see Section \ref{subsec:dataa-z}). We categorize the challenges facing domain-specific data applications and problems in terms of the following major areas: 
\begin{itemize}
\item Challenges in data/business understanding, 
\item Challenges in mathematical and statistical foundations,
\item Challenges in X-analytics and data/knowledge engineering, 
\item Challenges in data quality and social issues, 
\item Challenges in data value, impact and usability, 
\item Challenges in data-to-decision and actions. 
\end{itemize}

X-analytics and data/knowledge engineering encompass many specific research issues that have not been addressed properly; for example: 
\begin{itemize}
\item Behavior and event processing, 
\item Data storage and management systems, 
\item Data quality enhancement, 
\item Data modeling, learning and mining, 
\item Deep analytics, learning and discovery, 
\item Simulation and experimental design, 
\item High-performance processing and analytics, 
\item Analytics and computing architectures and infrastructure, 
\item Networking, communication and interoperation. 
\end{itemize}


\subsection{Course Framework of Data Science}
\label{subsec:courseframework}

The goal of data science and analytics education is to train and generate the data and analytics knowledge and proficiency required to manage the capability and capacity gaps in the creation of a data science profession \cite{Walker15,Manieri15-1}, and to achieve the goals of data science innovation and the data economy. Accordingly, different levels of education and training are necessary, from attending public courses, corporate training, and undergraduate courses, to joining a master of data science and/or PhD in data science program.

\emph{Public courses} are designed for the general community, to lift their understanding, skills, profession and specialism in data science through multi-level short courses. They range from basic courses to intermediary and advanced courses. The knowledge map consists of such components as data science, data mining, machine learning, statistics, data management, computing, programming, system analysis and design, and modules related to case studies, hands-on practices, project management, communication, and decision support.

\emph{Corporate training} and workshops are customized to upgrade and foster corporate thinking, knowledge, capability and practices for entire enterprise innovation and raising productivity. This involves offering courses and workshops  for the  workforce, from senior corporate executives to business owners, business analysts, data modelers, data scientists, data engineers, and deployment and enterprise strategists. Such courses cover scope and topics such as data science, data engineering, analytics science, decision science, data and analytics software engineering, project management, communications, and case management.

\emph{Undergraduate courses} may be offered on either a general data science basis that focuses on building data science foundations and data and analytics computing, or specific areas such as data engineering, predictive modeling, and visualization. Double degrees or majors may be offered to train professionals who will gain knowledge and abilities across disciplines such as business and analytics, or statistics and computing. 
 
\emph{Master of data science and analytics} aims to train specialists and foster the talent of those who have the capacity to conduct a deep understanding of data and  undertake analytics tasks in data mining, knowledge discovery and machine learning-based advanced analytics. Interdisciplinary experts may be trained from those who have a solid foundation in statistics, business, social science or other specific disciplines and are able to integrate data-driven exploration technologies with  disciplinary expertise and techniques. A critical area in which data science and analytics should be incorporated  is the classic master of business administration course. This is where new generation business leaders can be trained for the  new economy and a global view of economic growth.

\emph{PhD in data science and analytics} aims to train high level talent and specialists who have independent thinking, leadership, research, innovation and better practices for theoretical innovation to manage the significant knowledge and capability gaps, and for substantial economic innovation and raising productivity. Interdisciplinary research is encouraged to train leaders who have a systematic and strategic understanding of the what, how and why of data and economic innovation.              

\begin{figure*}[ht!]
\centerline{\includegraphics[width=0.85\textwidth]{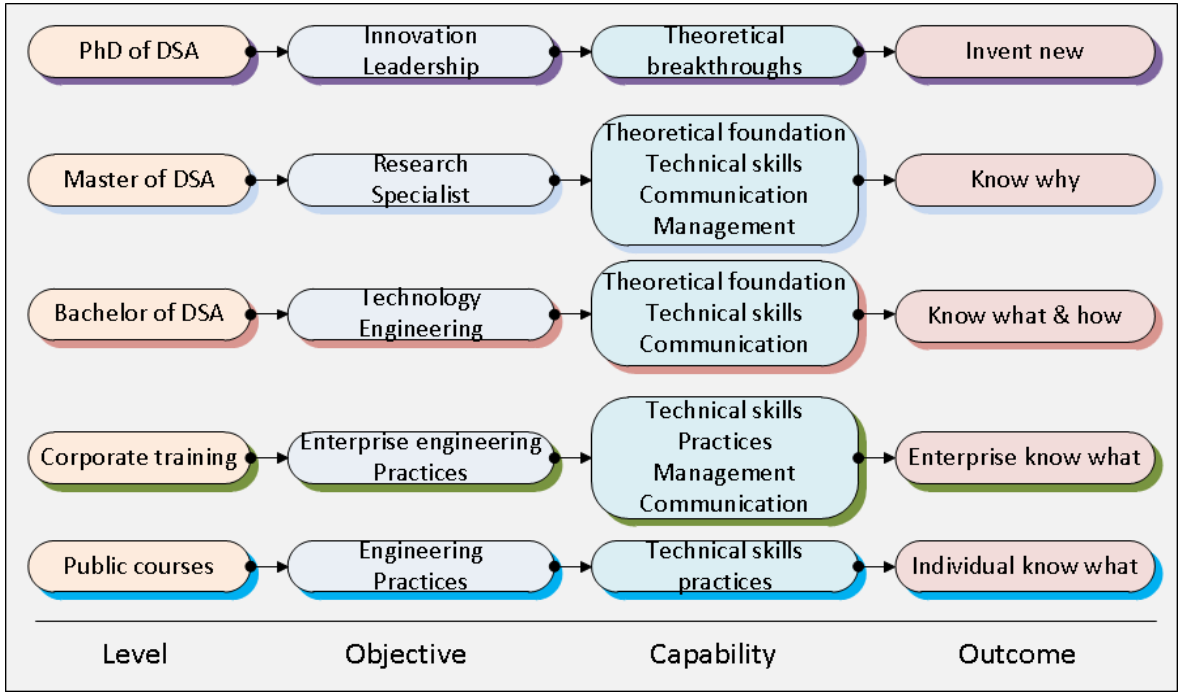}}
\caption{Data science course framework.}
\label{fig:12}
\end{figure*}

Fig. \ref{fig:12} shows the level, objective, capability set and outcomes of hierarchical data science and analytics education and training.

\section{Data Science as A New Science}
\label{sec:concepts}

So, what makes data science a new science? In this section, we discuss \textit{data A-Z}, which may be used to capture every aspect of data science to form a data science ontological system, \textit{the concept of data science}, which is built on the above discussions about the features and disciplinary development of data science, and \textit{the future of data science}. 

\subsection{Data A-Z}
\label{subsec:dataa-z}

In the big data community, multiple Vs are typically used to describe what constitutes big data, i.e., the characteristics, challenges and opportunities of big data. They include the volume (size), velocity (speed), variety (diversity), veracity (quality and trust), value (insight), visualization, and variability (formality) of data. 

In fact, these big Vs cannot describe a complete picture of big data, and  cannot capture the field of data science. Therefore, it is very valuable to build a \emph{data A-Z dictionary} to represent and capture the intrinsic comprehensive but diverse aspects, characteristics, challenges, domains, tasks, processes, purposes, applications and outcomes of, or related to, data. To this end, we list a sample sequence of data science keywords:

\begin{eqnarray}
Actionability/Adaptation, Behavior/Boosting, \nonumber \\
Causality/Change, Dimensionality/Divergence, \nonumber \\
Embedding/Ethics, Fusion/Forecasting, \nonumber \\
Governance/Generalization, Heterogeneity/ \nonumber \\
Hashing, Integrity/Inference, Join/Jungle, \nonumber \\
Kernelization/Knowledge, Linkage/Learning, \nonumber \\
Metrology/Migration, Normalization/Novelty, \nonumber \\
Optimization/Outlier, Privacy/Provenance, \nonumber \\
Quality/Quantity, Relation/Regularization, \nonumber \\
Scalability/Sparsity, Transformation/Transfer, \nonumber \\
Utility/Uncertainty, Variation/Visibility, \nonumber \\
Wrangling/Weighting, X-analytics/ \nonumber \\
X-informatics, Yield/Yipit, Zettabyte/Zenit. 
\end{eqnarray}

It is notable that such a data A-Z ontology  probably covers most of the topics of interest to major data science communities.
The exercise of constructing data A-Z can substantially deepen and broaden the understanding of intrinsic data characteristics, complexities, challenges, prospects and opportunities \cite{Geczy14}.

\subsection{What Is Data Science}
\label{subsec:ds}

Generally speaking, \textit{data science is the science of data}, which concerns the study of data. There are different ways to define what data science is: it may be object-focused, process-based, and/or discipline-oriented \cite{Cao16ds}. 
\begin{definition}[Data Science]
From the \textit{process} perspective, data science is a systematic approach to `think with wisdom', `understand domain', `manage data', `compute with data', `mine on knowledge', `communicate with stakeholders', `deliver products', and `act on insights'. 
\end{definition}

A \textit{process-based data science formula} is accordingly given below: 
\begin{eqnarray}
data~science = think + understand + manage + \nonumber \\ compute + mine + 
 communicate + deliver + act 
\end{eqnarray}

In contrast, \textit{data analytics understands data and its underlying business, discovers knowledge, delivers actionable insights, and enable decision-making}. From this perspective, we can say that analytics is a keystone of data science.

From the \textit{disciplinary} perspective, \textit{data science is a new interdisciplinary field} in which to study data and its domain in terms of a data-to-knowledge-to-wisdom thinking for generating data products \cite{Cao16ds}. Data science integrates \textit{traditionally data-oriented disciplines} such as statistics, informatics and computing with \textit{traditionally data-independent fields} such as communication, management and sociology. 

\subsection{The Future of Data Science}
\label{subsec:dsfuture}

It is difficult at this very early stage of data science to predict specific future data science innovation and research, thus the next-generation data science will need to address the unknown space that is currently invisible to existing science and create new data products. We will need to:
\begin{itemize}
\item Deepen our \textit{understanding of data invisibility} (i.e., \textit{invisible data characteristics}) in the hidden and blind spaces (Spaces B and D in Fig. 1 \cite{Cao16ds}), to understand their X-complexities (see \cite{Cao16ds}) and X-intelligence (see \cite{Cao16ds}), and strengthen our capabilities;
\item Invent \textit{new data representation capabilities}, including designs, structures, schemas and algorithms to make invisible data in Spaces B and D in Fig. 1 \cite{Cao16ds} more visible and explicit;
\item Create \textit{new analytical and learning capabilities}, including original theories, algorithms and models, to disclose the unknown knowledge in unknown Space D in Fig. 1 \cite{Cao16ds};
\item Build new intelligent systems and services, including corporate and Internet-based collaborative platforms and services, to support collaborative and collective exploration of invisible and unknown challenges in the fully unknown space D in Fig. 1 \cite{Cao16ds}.
\item Train a generation of qualified data science professionals in data literacy, thinking, competency, consciousness and cognitive intelligence to work on the above data science agenda.
\end{itemize}

\section{Pitfalls in Data Science}
\label{sec:pitfall}

A typical feature of data science being at this very early stage is that different and sometimes contradictory views appear in various communities. It is essential to share and discuss the myths and reality \cite{Jagadish15}, memes \cite{Donoho15}, and pitfalls to ensure the healthy development of the field. Based on our observations about the relevant communities, and our experience and lessons learned in conducting data science and analytics research, education and services, we list the following myths and pitfalls for discussion. 

\subsection{About Data Science Concepts}
\label{subsec:dsc}

Typically, data science has been defined in terms of specific disciplinary foundations, principles, goals, inputs, algorithms and models, processes, tools, outputs, applications, and/or professions. Often, a fragmented statement is given, which may cause debate and result in the phenomenon of ``how does a bind person recognize an elephant?'' In this section, we discuss some common arguments and observations.
\begin{itemize}
\item Data science is statistics \cite{Broman13,Diggle15}; ``why do we need data science when we've had statistics for centuries'' \cite{Wladawsky14}? How does data science really differ from statistics \cite{Donoho15}? (Comments: Data science provides systematic, holistic and multi-disciplinary solutions for learning explicit and implicit insights and intelligence from complex and large-scale data and generates evidence or indicators from data by undertaking diagnostic, descriptive, predictive and/or prescriptive analytics, in addition to supporting other tasks on data such as computing and management.)
\item Why do we need data science when information science and data engineering have been explored for many years? (Comments: Consider the issues faced in related areas by the enormity of the task and the parallel example of enabling a blind person to recognize an animal as large as an elephant.  Information science and data engineering alone cannot achieve this. Other aspects may be learned from the discussion about greater or fewer statistics; more in \cite{Chambers93}.)
\item I have been doing data analysis for dozens of years; data science has nothing new to offer me. (Comments: Classic data analysis and technologies focus mostly on explicit observation analysis and hypothesis testing on small and simpler data.)
\item Is data science old wine in a new bottle? What are the new grand challenges foregrounded by data science? (Comments: The analysis of the gaps between existing developments and the potential of data science (see Fig. \ref{fig:four}) shows that many opportunities can be found to fill the theoretical gaps when data complexities extend significantly beyond the level that can be handled by the state-of-the-art theories and systems, e.g., classic statistical and analytical theories and systems were not designed to handle the non-IIDness \cite{noniid14} in complex real-life systems.)
\item Data science mixes statistics, data engineering and computing, and does not contribute to breakthrough research. (Comments: Data science attracts attention because of the significant complexities in handling complex real-world data, applications and problems that cannot be addressed well by existing statistics, data engineering and computing theories and systems. This drives significant innovation and produces unique opportunities to generate breakthrough theories.)
\item Data science is also referred to as data analytics and big data \cite{Anderson14}. (Comments: This confuses the main objectives, features, scopes of the three concepts and areas. Data science needs to be clearly distinguished from both data analytics and big data.)
\item Other definitions ascribed to data science are that it is big data discovery \cite{Davenport12}, prediction \cite{Dhar13}, or the combination of principle and process with technique \cite{Provost13}.
\end{itemize}

It is also worth noting attention that the terms ``big data'', ``data science'' and ``advanced analytics'' are often extensively misused, over-used or improperly used by diverse communities and for various purposes, particularly given the influence of media hype and buzz. A large proportion of Google searches on these  keywords returns results that are irrelevant to their intrinsic semantics and scope, or simply repeat familiar arguments about the needs of data science and existing phenomena. In many  such findings \cite{Brown,Stanton12,Chawla14,Faris11,Anya15,Dierick15,Priebe15,Kanter15,Kirkpatrick15,Clancy14,Stevens14,Miller13,
Pal15,Moraes15,Horton15,Loukides12,Neil13,Hand15,Hazena14,Manieri15-2,Cuzzocrea13,Casey15,Siart15,Gold13,Gupta15}, big data is described as being simple, data science has nothing to do with the science of data, and advanced analytics is  the same as classic data analysis and information processing. There is a lack of deep thinking and exploration of why, what and how these new terms should be defined, developed and applied. 

The above observations strongly illustrate that data science is still in a very early stage. They also justify the urgent need to develop sound terminology, standards, a code of conduct, statement and definitions, theoretical frameworks, and better practices that will exemplify typical data science professional practices and profiles.  

\subsection{About Data Volume}
\label{subsec:dv}
\begin{itemize}
\item What makes data ``big''? (Comments: It is usually not the volume but the complexities (as discussed in \cite{Cao16ds,Metasynthetic15}) and large values that make data big.)
\item Why is the bigness of data important? (Comments: The bigness (referring to data science complexities) of data heralds new opportunities for theoretical, technological, practical, economic and other development or revolution.)
\item Big data refers to massive volumes of data. (Comments: Here, ``big'' refers mainly to significant data complexities. From the volume perspective, a data set is big when the size of the data itself becomes a quintessential part of the problem.)
\item Data science is big data analytics. (Comments: Data science is a comprehensive field centered on manipulating data complexities and extracting intelligence, in which data can be big or small and analytics is a core component and task.)
\item I do not have big data so I cannot do big data research. (Comments: Most researchers and practitioners do not have sizeable amounts of data and do not have access to big infrastructure either. However, significant research opportunities still exist to create fundamentally new theories and tools to address respective X-complexities and X-intelligence.)
\item The data I can find is small and too simple to be explored. (Comments: While scale is a critical issue in data science, small data, which is widely available, may still incorporate interesting data complexities that have not been well addressed.  Often, we see experimental data, which is usually small, neat and clean. Observational data from real business is live, complex, large and frequently messy.)
\item I am collecting data from all sources in order to conduct big data analytics. (Comments: Only relevant data is required to achieve a specific analytical goal.)
\item It is better to have too much data than too little. (Comments: While more data generally tends to present more opportunities, the data amount needs to be relevant to the data needed and the data manipulation goals. Whether bigger is better depends on many aspects.)
\end{itemize}

\subsection{About Data Infrastructure}
\label{subsec:di}
\begin{itemize}
\item I do not have big infrastructure, so I cannot do big data research. (Comments: While big infrastructure is useful or necessary for some big data tasks, theoretical research on significant challenges may not require big infrastructure.)
\item My organization will purchase a high performance computer to support big data analytics (Comments: Many big data analytics tasks can be done without a high performance computer. It is also essential to differentiate between distributed/parallel computing and high performance computing.)
\end{itemize}

\subsection{About Analytics}
\label{subsec:ana}
\begin{itemize}
\item Thinking data-analytically is crucial for data science. (Comments: Data-analytic thinking is not only important for a specific problem-solving, but is essential for obtaining a systematic solution and for a data-rich organization. Converting an organization to think data analytically is a critical competitive advantage in the data era.)
\item The task of an analyst is mainly to develop common task frameworks and conduct inference \cite{Breiman01} from the particular to the general. (Comments: Analytics in the real world is often specific. Focusing on certain common task frameworks may trigger incomplete or even misleading outcomes. As discussed in Section \ref{subsec:ds}, an analyst may take other roles, e.g., predictive modeling is typically problem-specific.)
\item I only trust the quality of models built in commercial analytical tools. (Comments: Such tools may produce misleading or even incorrect outcomes if the assumption of their theoretical foundation does not fit the data, e.g., if they only suit imbalanced data, normal distribution-based data, or IID data.)
\item Most published models and algorithms and their experimental outcomes are not repeatable. (Comments: Such works seem to be more hand-crafted rather than manufactured. Repeatability, reproducibility, open data and data sharing are critical to the healthy field development.)  
\item I want to do big data analytics, can you tell me which algorithms and program language I should learn? (Comments: Public survey outcomes (see examples in \cite{cao16-2}) give responses to such questions. Which algorithms, language and platform should be chosen also depends on  organizational maturity and needs. For long-term purposes, big data analytics is about building competencies rather than specific functions).
\item My organization's data is private and thus you cannot be involved in our analytics. (Comments: Private data can still be explored by external parties by implementing proper privacy protection and setting up appropriate policies for onsite exploration.) 
\item Let me (an analyst) show you (business people) some of my findings which are statistically significant. (Comments: As domain-driven data mining \cite{dddm10} shows, many outcomes are often statistically significant but are not actionable. An evaluation of those findings needs to be conducted to discover what business impact  \cite{cao08-3} might be generated if the findings they may generate are operationalized.)
\item Strange, why can I not understand and interpret the outcomes? (Comments: This may be because the problem  has been misstated, the model may be invalid for the data, or the data used is not relevant or correct.)
\item Your outcomes are too empirical without theoretical proof and foundation. (Comments: While it would be ideal if  questions about the outcomes could be addressed from  theoretical, optimization and evaluation perspectives, real-life complex data analytics may often be more exploratory and it may initially be difficult to optimize empirical performance.)
\item My analysis shows what you delivered is not the best for our organization. (Comments: It may be challenging to claim ``the best'' when a variety of models, workflows and  data features are used in analytics. It is not unusual for analysts to obtain different or  contradictory outcomes on the same data as a result of the application of different theories, settings and models. It may turn out to be a very challenging job to find a solid model that perfectly and stably fits the invisible aspect of data characteristics. It is important to appropriately check the relevance and validity of the data, models, frameworks and workflows available and used. Doing the right thing at the right time for the right purpose is a very difficult task when attempting to understand complex real-life data and problems.)
\item Can your model address all of my business problems? (Comments: Different models are often are required to address diverse business problems, as  a single model cannot handle a problem sufficiently well.)
\item This model is very advanced with solid theoretical foundation, let us try it in your business. (Comments: While having solid scientific understanding of a model is important, it is data-driven discovery  may better capture the actual data characteristics in real-life problem solving. A model may be improperly used without a deep understanding of model and data suitability. Combining data driven approaches with model driven approaches may be more practical.) 
\item My analytical reports consist of lots of figures and tables that summarize the data mining outcomes, but my boss seems not so interested in them. (Comments: Analytics is not just about producing meaningful analytical outcomes and reports; rather, it concerns insights, recommendations and communication with upper management for decision-making and action.)   
\item It is better to have advanced rather than simple models. (Comments: Generally, simpler is better. The key to deploying a model is to fit the model to the data while following the same assumption taken by the model.)
\item We just tuned the models last month, but again they do not work well. (Comments: Monitoring a model's performance  by watching the dynamics and significant change that may take place in the data and business is critical. Real-time analytics requires adaptive and automated re-learning and adjustment.) 
\item I designed the model, so I trust the outcomes. (Comments: The reproducibility of model outcomes relies on many factors. A model that is properly constructed may fall short in other aspects such as data leakage, overfitting, insufficient data cleaning, and poor understanding of data characteristics and business. Similarly, a lack of communication with the business may cause serious problems in the quality of  the outcome.)
\item Data science and analytics projects are just other kinds of IT projects. (Comments: While data projects share many similar aspects to mainstream IT projects, certain distinctive features in data, the manipulation process, delivery, and especially the exploratory nature of data science and analytics projects require different strategies, procedures and treatments. Data science projects are more exploratory, ad hoc, decision-oriented and intelligence-driven.)
\end{itemize}

\subsection{About Capabilities and Roles}
\label{subsec:cap}
\begin{itemize}
\item I am a data scientist. (Comments: Lately, it seems that everyone has suddenly become a data scientist. Most data scientists simply conduct normal data engineering and descriptive analytics. Do not expect omnipotence from data scientists.)
\item ``A human investigative journalist can look at the facts, identify what's wrong with the situation, uncover the truth, and write a story that places the facts in context. A computer can't.'' \cite{Kirkpatrick15} (Comments: The success of AlphaGo  \cite{deepmind} may show the potential that a data science-enabled computer has to undertake a large proportion of the job a journalist does.)
\item My organization wants to do big data analytics, can you recommend some of your PhD graduates to us? (Comments: While data science and advanced analytics tasks usually benefit from the input of PhDs,  an organization requires different roles and competencies according to the maturity level of the analytics and the organization.)
\item Our data science team consists of a group of data scientists. (Comments: An effective data science team may consist of statisticians, programmers, physicists, artists, social scientists, decision-makers, or even entrepreneurs.)
\item A data scientist is a statistical programmer. (Comments: In addition to the core skills of coding and statistics, a data scientist needs to handle many other matters; see discussions in \cite{cao16-2}.)
\end{itemize}

\subsection{Other Matters}
In addition to the above aspects, there are other matters that require careful consideration in conducting data science and analytics. We list some here.
\begin{itemize}
\item Garbage in, garbage out. (Comments: The quality of data determines the quality of output.)
\item More complex data, a more advanced model, and better outcomes.  (Comments: Good data does not necessarily lead to good outcomes; A good model also does not guarantee good outcomes.) 
\item More general models, better applicability. (Comments: General models may lead to weaker outcomes on a specific problem. It is not reasonable or practical to expect a single tool for all tasks.)
\item More frequent patterns, more interesting. (Comments: It has been shown that  frequent patterns mined by existing theories are generally not useful and actionable.) 
\item We're interested in outcomes, not theories. (Comments: Actionable outcomes may need to satisfy both technical and business significance \cite{dddm10}.)
\item The goal of analytics is to support decision-making actions, not just to present outcomes about data understanding and analytical results. (Comments: This addresses the need for actionable knowledge delivery \cite{Cao10dddm} to recommend actions from data analytics for decision-support.)
\item Whatever you do, you can at least get some values. (Comments: This is true, but it may be risky or misleading. Informed data manipulation and analytics requires a foundation for interpreting why the outcomes look the way they do.)
\item Many end users are investing in big data infrastructure without project management. (Comments: Do not rush into data infrastructure investment without a solid strategic plan of your data science initiatives, which requires the identification of business needs and requirements, the definition of reasonable objectives, the specification of timelines, and the allocation of resources.)
\item Pushing data science forward without suitable talent. (Comments: On one hand, you should not simply wait for the right candidate to come along, but should, actively plan and specify the skills needed for your organization's initiatives and assemble a team according to the skill-sets required. On the other hand, getting the right people on board is critical, as data science is essentially about intelligence and talent.)
\item No culture for converting data science insights into actionable outcomes. (Comments: This may be common in business intelligence and technically focused teams. Fostering a data science-friendly culture requires  a top-down approach driven by business needs, and making data-driven decisions that  enable data science specialists and project managers to be part of the business process, and to conduct change management.)
\item Correct evaluation of outcomes. (Comments: This goes far beyond such technical metrics as Area Under the ROC Curve and Normalized Mutual Information. Business performance after adopting the recommended outcomes needs to be evaluated \cite{Cao10dddm}. For example, recent work on high utility analysis \cite{yin12} and high impact behavior analysis \cite{cao08-3} study how business performance can be taken into data modeling and evaluation account. Solutions that lack business viability are not actionable.)   
\item Apply a model in a consistent way. (Comments: It is essential to understand the hypothesis behind a model and to apply a model consistent with its hypothesis.)
\item Overthinking and overusing models. (Comments: All models and methods are specific to certain hypotheses and scenarios. No models are universal and sufficiently ``advanced'' to suit everything. Do not assume that if the data is tortured long enough, it will confess to anything.) 
\item Know nothing about the data before applying a model. (Comments: Data understanding is a must-do step before a model is applied.)
\item Analyze data for the sake of analysis only. (Comments: This involves the common bad practice of overusing analytics.)
\item What makes an insight (knowledge) actionable? (Comments: This is dependent on not only the statistical and practical values of the insight, but also predictive power and business impact.)
\item Do not assume the data you are given is perfect. (Comments: Data quality forms the basis of obtaining good models, outcomes and decisions. Poor quality  data, the same as poor quality models, can lead to misleading or damaging decisions. Real-life data often contains imperfect features such as incompleteness, uncertainty, bias, rareness, imbalance and non-IIDness.)
\end{itemize}

\section{Conclusions}
\label{sec:concl}
In the era of data science, big data and advanced analytics, numerous debates have emerged from a wide range of backgrounds, domains, areas and perspectives and for diversified reasons and purposes. It is difficult but critical to explore the nature of data science. To do so, a fundamental perspective is to explore the intrinsic characteristics, challenges, working mechanisms, and dynamics of data and the science about data.

As part of our comprehensive review of data science \cite{cao16-1,cao16-2,Cao16ds}, the discussions about the nature and pitfalls of data science in this work will hopefully stimulate deep and intrinsic discussions about what makes data science a new science, and what makes data science valuable for research, innovation, the economy, services and professionals.

\section*{Acknowledgment}
This work is partially sponsored by the Australian Research Council Discovery Grant (DP130102691).

 \ifCLASSOPTIONcaptionsoff
  \newpage
\fi


\bibliographystyle{abbrv} 
\bibliography{ds-ieeeis.17}

\end{document}